\documentclass[twocolumn]{aastex631}

\usepackage{booktabs}
\usepackage{caption}
\usepackage{subcaption}
\usepackage{textcomp}
\usepackage{tipa}
\usepackage{float}

\newcommand{\rh}{$r_{\text{h}}$}

\newcommand{\Msun}{$\textup{M}_\odot$}

\newcommand{\met}{[Fe/H]}

\newcommand{\peg}{Peg VII}

\shorttitle{Pegasus VII}
\shortauthors{Smith et al.}

\graphicspath{{./}{figures/}}

\begin{document}

\title{Deep in the Fields of the Andromeda Halo: \\Discovery of the Pegasus VII dwarf galaxy in UNIONS}

\correspondingauthor{Simon E.\,T. Smith}
\email{simonsmith@uvic.ca}

\author[0000-0002-6946-8280]{Simon E.\,T. Smith}
\affiliation{Department of Physics and Astronomy, University of Victoria, Victoria, BC, V8P 1A1, Canada}

\author{Alan W. McConnachie}
\affiliation{NRC Herzberg Astronomy and Astrophysics, 5071 West Saanich Road, Victoria, BC, V9E 2E7, Canada}
\affiliation{Department of Physics and Astronomy, University of Victoria, Victoria, BC, V8P 1A1, Canada}

\author{Stephen Gwyn}
\affiliation{NRC Herzberg Astronomy and Astrophysics, 5071 West Saanich Road, Victoria, BC, V9E 2E7, Canada}

\author{Christian R. Hayes}
\affiliation{Space Telescope Science Institute, 3700 San Martin Drive, Baltimore, MD 21218, USA}

\author{Massimiliano Gatto}
\affiliation{INAF - Osservatorio Astronomico di Capodimonte, Via Moiariello 16, I-80131 Naples, Italy}

\author{Ken Chambers}
\affiliation{Institute for Astronomy, University of Hawaii, 2680 Woodlawn Drive, Honolulu HI 96822}

\author{Jean-Charles Cuillandre}
\affiliation{AIM, CEA, CNRS, Universit\'e Paris-Saclay, Universit\'e Paris, F-91191 Gif-sur-Yvette, France}

\author{Michael J. Hudson}
\affiliation{Department of Physics and Astronomy, University of Waterloo, 200 University Ave W, Waterloo, ON N2L 3G1, Canada}
\affiliation{Waterloo Centre for Astrophysics, University of Waterloo, 200 University Ave W, Waterloo, ON N2L 3G1, Canada}
\affiliation{Perimeter Institute for Theoretical Physics, 31 Caroline St. North, Waterloo, ON N2L 2Y5, Canada}

\author{Eugene Magnier}
\affiliation{Institute for Astronomy, University of Hawaii, 2680 Woodlawn Drive, Honolulu HI 96822}

\author{Nicolas Martin}
\affiliation{Universit\'e de Strasbourg, CNRS, Observatoire Astronomique de Strasbourg, UMR 7550, F-67000 Strasbourg, France}
\affiliation{Max-Planck-Institut f\"ur Astronomie, K\"onigstuhl 17, D-69117, Heidelberg, Germany}

\author{Julio Navarro}
\affiliation{Department of Physics and Astronomy, University of Victoria, Victoria, BC, V8P 1A1, Canada}

\begin{abstract}
    We present the newly discovered dwarf galaxy Pegasus VII (\peg), a member of the M31 sub-group which has been uncovered in the $ri$ photometric catalogs from the Ultraviolet Near-Infrared Optical Northern Survey and confirmed with follow-up imaging from both the Canada-France-Hawaii Telescope and the Gemini-North Telescope. This system has an absolute $V$-band magnitude of $-5.7$\,$\pm$\,0.2\,mag and a physical half-light radius of 177$^{+36}_{-34}$\,pc, which is characteristic of dynamically-confirmed Milky Way satellite dwarf galaxies and about 5 times more extended than the most extended M31 globular clusters. \peg\ lies at a three-dimensional separation from M31 of 331$^{+15}_{-4}$\,kpc and 
    a significant elongation ($\epsilon \sim 0.5$) towards the projected direction of M31 could be indicative of a past tidal interaction, but additional investigation into the orbit, star formation history, and whether any gas remains in the galaxy is needed to better understand the evolution of \peg.
\end{abstract}

\section{Introduction} \label{sec:intro}

Every newly discovered dwarf galaxy is a unique laboratory to study the interplay between chemical enrichment \citep[e.g.,][]{Ji16-ret2, Ji16-tuc2, Waller2023} and stellar feedback \citep[e.g.,][]{Bovill09, Bullock17}, the effect of reionization \citep[e.g.,][]{Bullock00, Brown2012, Brown2014}, and the impact of tidal interactions on galactic structure \citep[e.g.,][]{Weisz2011, Weisz2014, Errani2021, Errani22}.

From a dwarf galaxy satellite population standpoint, the M31 system provides an excellent prospect for study, because our relative proximity and external perspective offer the possibility of a spatially complete mapping of the M31 halo. The total membership and arrangement of M31 satellite galaxies offers a potential probe of the particle nature and small-scale distribution of dark matter \citep[e.g.,][]{Wheeler15, Nadler21, Silverman23}, testing dark matter theories that were initially invoked to explain structures on the largest cosmological scales \citep[][]{Lovell12, Bullock17, Sales2022}. This is highly complementary to similar studies in the Milky Way halo which naturally uncover satellites with less intrinsic luminosity that are closer, but which will always be spatially limited due to the presence of the Milky Way disk.

The most productive search for M31 substructures have been the Pan-Andromeda Archaeological Survey \citep[PAndAS;][]{McConnachie2009}, where 19 new satellite galaxies have been discovered \citep[e.g.,][]{Martin2009, Richardson2011}, along with $\sim$100 globular clusters \citep[][]{Huxor2014}, and a wide variety of low surface brightness features \citep[e.g.,][]{Martin2014-fos, Ibata2014, McConnachie2018}. However, PAndAS primarily surveyed the inner 150\,kpc of the M31 halo.

\begin{figure*}
    \centering
    \includegraphics[width=0.93\linewidth]{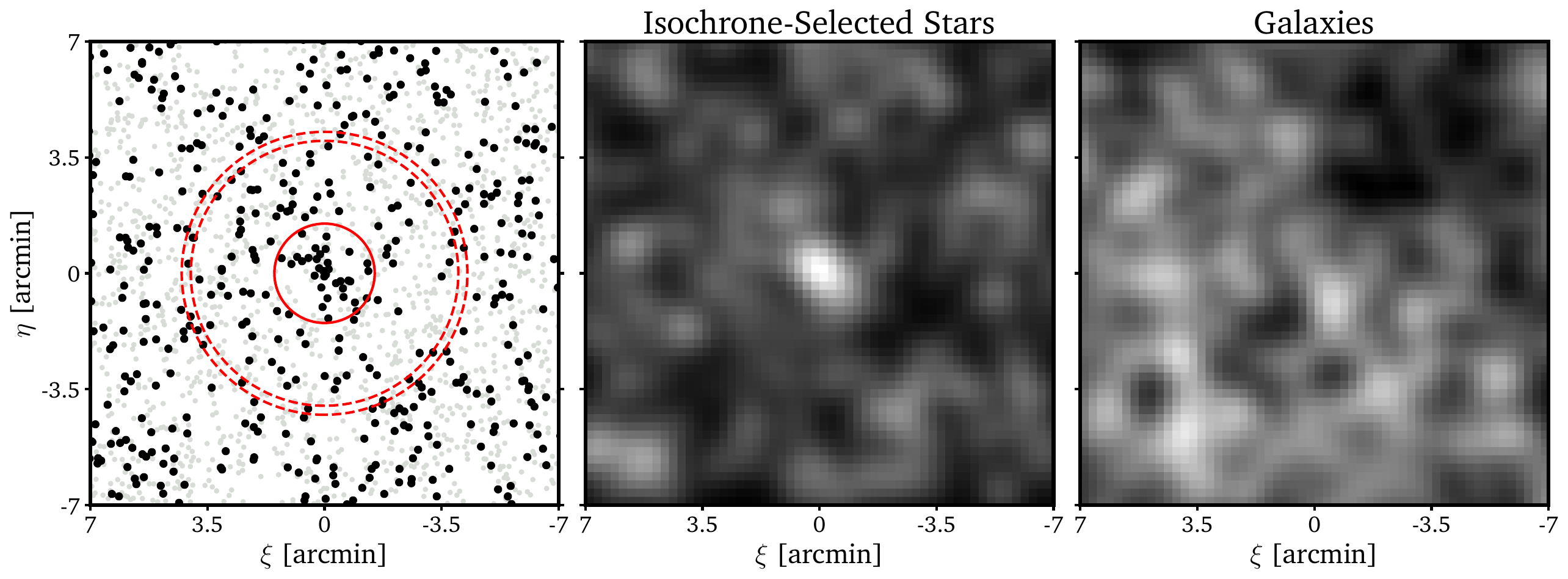}
    \includegraphics[width=0.62\linewidth]{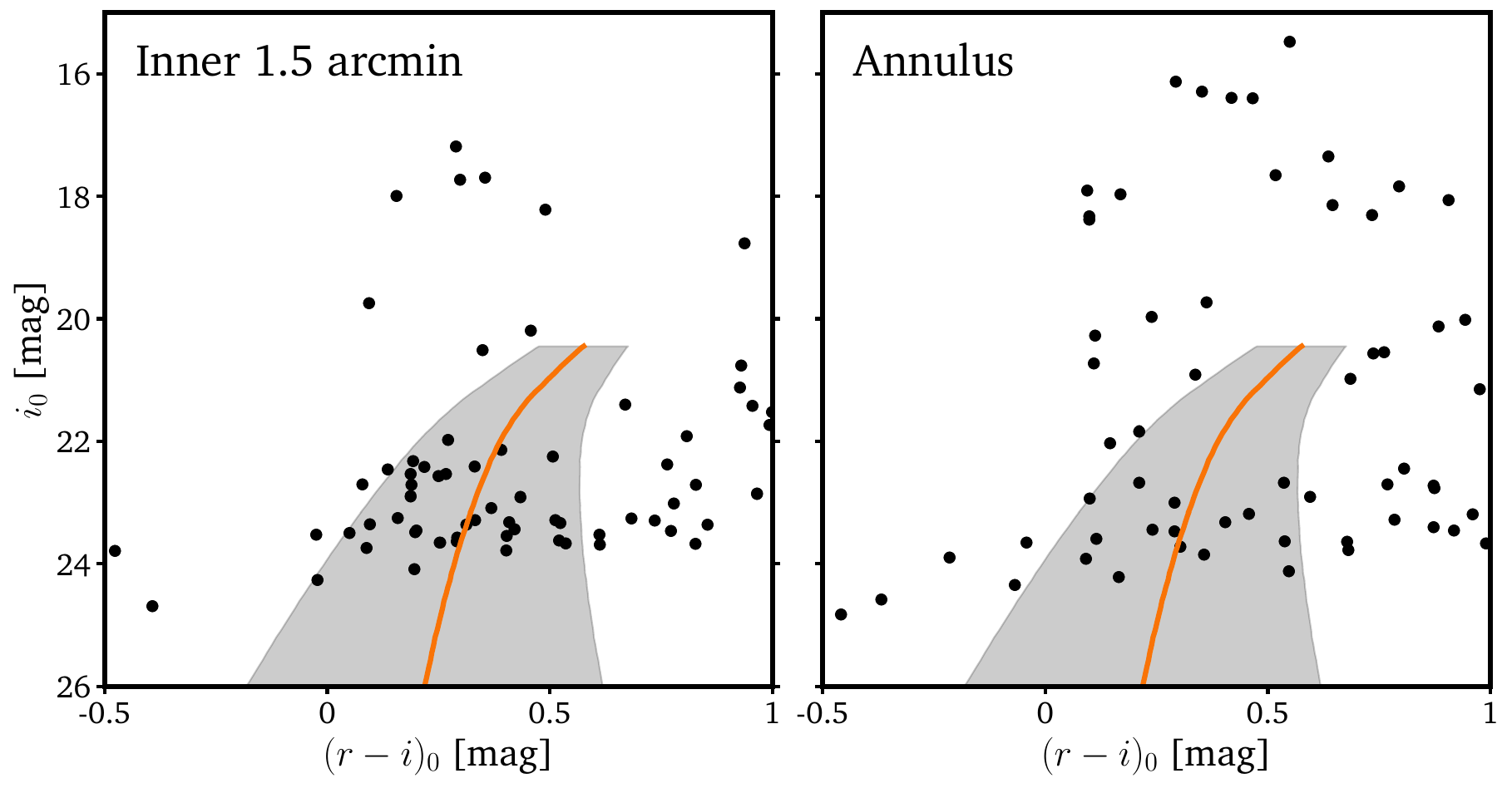}
    \caption{A series of plots showing the tentative detection of a candidate stellar overdensity (\peg) in the UNIONS photometric catalogs.
    {\it Upper Left}: Spatial distribution of stellar sources in 7\arcmin\,$\times$\,7\arcmin\ region about the candidate. Black points are stars that fall within the shaded isochrone selection region in the Color-Magnitude Diagrams (CMDs) in the right panels. The inner solid red circle has a radius of 1.5\arcmin\ and the dashed red annulus is equal area with the inner circle, having an inner radius of 4\arcmin.
    {\it Upper Middle}: Smoothed spatial distribution of isochrone-selected stars.
    {\it Upper Right}: Smoothed spatial distribution of isochrone-selected all galaxies detected in UNIONS in the same field of view as the upper middle panel. No overdensity is visible at the location of the candidate dwarf galaxy.
    {\it Lower Left}: CMD of extinction-corrected $ri$ UNIONS photometry. All sources within 1.5\arcmin\ are shown. An old (12\,Gyr), metal-poor ([Fe/H] = $-2$) isochrone is overlaid, shifted to a distance of 700\,kpc. The grey shaded region shows the isochrone selection of potential member stars.
    {\it Lower Right}: Same as {\it Lower Left} plot, but for stars in the annulus from the {\it Upper Left} plot, showing one example of the field contamination on the target CMD.
    }
    \label{fig:unions}
\end{figure*}

Satellites in the outer reaches of the Andromeda system \citep[beyond 150\,kpc in projection; e.g.,][]{Slater2011, Martin2013-per, Martin2013-2new, MDelgado22, Collins22, McQuinn23-PegW} were uncovered through the systematic exploration of wide-field surveys, namely the Sloan Digital Sky Survey \citep[SDSS;][]{Abazajian09}, the Panoramic Survey Telescope and Rapid Response System (Pan-STARRS) 3$\pi$ survey \citep[][]{Chambers16}, and the Dark Energy Spectroscopic Instrument (DESI) Legacy Imaging surveys \citep[][]{Dey2019}. These surveys cover vast swaths of the extragalactic skies in the M31 vicinity, albeit at generally shallower photometric limits the median 5$\sigma$ depths of $g \sim 26$ and $i \sim 24.8$ of PAndAS. New to the game of M31 satellite discovery is the deep, wide-field Ultraviolet Near-Infrared Optical Northern Survey (UNIONS; Gwyn et al. 2025 (in preparation), a 5000\,deg$^2$ $ugriz$ survey, part of which covers a several hundred square degree region of sky just south of M31 at depths less than 1\,mag shallower than PAndAS, making it the deepest available survey for exploring the far reaches of the M31 halo. The quality and breadth of UNIONS has already led to the identification of two new Milky Way satellites \citep[][]{Smith23, Smith24}.

In this work, we present the discovery of a faint dwarf galaxy candidate in the constellation of Pegasus. We adopt the name Pegasus VII (hereafter \peg) as the name for this newly discovered Local Group dwarf galaxy candidate, given that the recently discovered Pegasus W \citep{McQuinn23-PegW} is the sixth dwarf galaxy in the Local Group found in this constellation. \peg\ is faint ($M_V = -5.7$\,$\pm$\,0.2\,mag) and has been uncovered at a large physical separation (331$^{+15}_{-4}$\,kpc) from M31 in UNIONS.
The paper is arranged as follows:
Section \ref{sec:methods} describes detection in the UNIONS data set and subsequent follow-up imaging with the Gemini and CFHT telescopes to confirm the candidate, Section \ref{sec:res} presents derived quantities for \peg, and we conclude with a discussion of this galaxy's potential origins and the state of the M31 satellite population in Section \ref{sec:disc}.

\section{Data and Detection} \label{sec:methods}
\subsection{UNIONS}
UNIONS is a consortium of independent photometric surveys, namely the Canada-France Imaging Survey (CFIS, $ur$) at the Canada-France-Hawaii Telescope (CFHT), Pan-STARRS ($iz$), the Wide Imaging with Subaru HSC of the Euclid Sky (WISHES, $z$) program, and the Waterloo-Hawaii IfA $g$-band Survey (WHIGS, $g$) at the Subaru Telescope, that have been brought together under a single heading to produce combined $ugriz$ imaging. The specific photometric bands contributed by each observing program are indicated in italics. A full description of UNIONS is available in Gwyn et al. 2025 (in preparation).

While one major aim of UNIONS is to support the Euclid space mission \citep{Laureijs2011, EC2024} with ground-based imaging for photometric redshifts, UNIONS is an independent imaging survey which seeks to maximize the science returns of this deep, five-band photometric survey.
UNIONS is covering 5000\,deg$^2$ of extragalactic ($\delta > 30$\textdegree), northern ($|b| > 30$\textdegree) sky when completed, at depths roughly comparable to year one of the Legacy Survey of Space and Time \citep[LSST;][]{Ivezic2019} at the Rubin Observatory. There is no planned ground-based northern survey that will surpass UNIONS in depth and breadth in the coming decade, nor does the primary UNIONS survey footprint intersect with the currently planned LSST footprint, making UNIONS an integral resource for photometric exploration of the northern extragalactic skies for the foreseeable future.

\begin{figure}
    \centering
    \includegraphics[width=\linewidth]{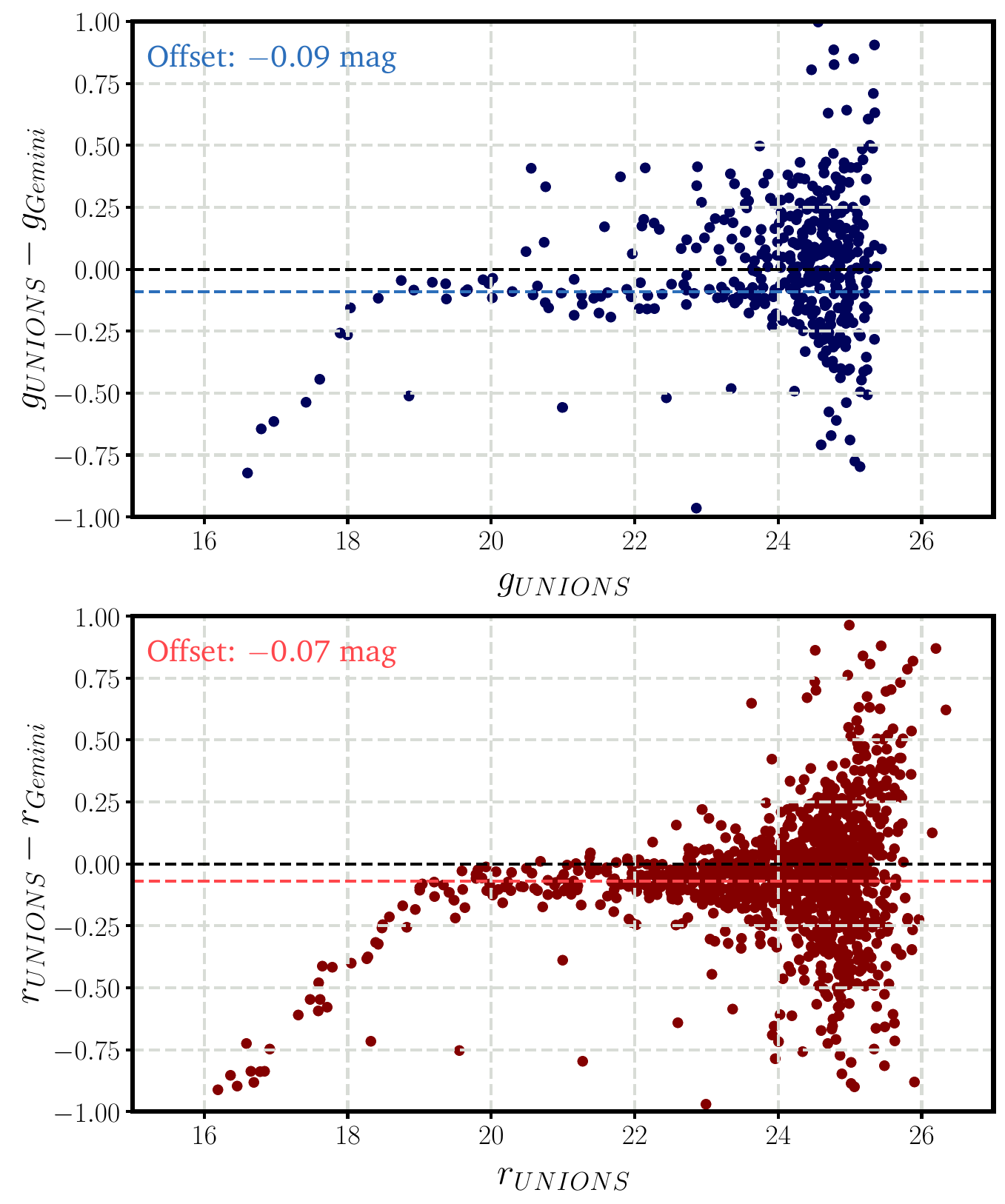}
    \caption{
    {\it Top}: $g_{\text{UNIONS}} - g_{\text{Gemini}}$ as a function of $g_{\text{UNIONS}}$. A constant offset of 0.09\,mag (blue dashed line) is needed to bring the Gemini photometry into agreement with the UNIONS photometry.
    {\it Bottom}: $r_{\text{UNIONS}} - r_{\text{Gemini}}$ as a function of $r_{\text{UNIONS}}$. A constant offset of 0.07\,mag (red dashed line) is needed to bring the Gemini photometry into agreement with the UNIONS photometry. In both panels, the downwards trend of the locus at bright magnitudes is due to saturation.
    }
    \label{fig:offsets}
\end{figure}

\subsection{Identification in UNIONS}
The initial detection of the \peg\ candidate system was carried out using the photometric catalogs derived from the CFIS $r$-band and Pan-STARRS $i$-band imaging, which currently have 5-sigma point sources depths of 24.9 and 24.0\,mag, respectively, over a shared area of roughly 4300\,deg$^2$. 

Star-galaxy separation is performed morphologically in the $r$-band to take advantage of its outstanding image quality (median diameter of the seeing disk for point sources is 0.69\arcsec). Galactic dust extinction corrections are applied using $E(B-V)$ values from \citet{Schlegel98} assuming the conversion factors of \citet{Schlafly11} for a reddening parameter of $R_V = 3.1$. The reddening for the CFHT $r$-band filter is not computed in \citet{Schlafly11}, so we adopt the conversion factor for the $r$-band filter on the Dark Energy Camera \citep{Flaugher15} as the wavelength range and transmission properties are nearly identical to those of the CFHT $r$-band.

\peg\ was identified during an ongoing search for Local Group dwarf galaxies in UNIONS that has previously produced the discoveries of Bo\"otes V (\citet{Smith23}, co-discovered by \citealt{Cerny23-6}) and Ursa Major III/UNIONS 1 \citep{Smith24}, with the latter being the faintest Milky Way satellite yet known, demonstrating the power of UNIONS for identifying Local Group satellites. The matched-filter based detection method scans for overdensities of old, metal-poor stars at heliocentric distances from 10--1000\,kpc; the full methodology is described in more detail in \citet{Smith23}.
\peg\ was flagged as a marginal overdensity of stars at distances of 700--1000\,kpc towards M31 in combined $ri$ UNIONS photometry, at a statistical significance slightly lower than that of several previously known M31 satellite dwarf galaxies that had been found in PAndAS and are also in the UNIONS footprint. \peg\ itself is located well outside of the PAndAS footprint. 

In Figure \ref{fig:unions} we show the spatial overdensity of stars, both as point sources and as a smoothed distribution, in the combined UNIONS $ri$ catalog that indicate the candidate dwarf galaxy. We plot a CMD of stars in a circular region with a radius of 1.5\arcmin\ centered on the detection centroid, with an old (12\,Gyr), metal-poor ([Fe/H] = $-2$) PARSEC isochrone \citep{Bressan12} shifted to a nominal distance of 700\,kpc. An additional CMD is shown with stars in an equal-area annulus centered on the detection centroid, whose inner radius is 4\arcmin, demonstrating the CMD contamination from foreground and background sources. On the CMD, a shaded grey region shows our rough selection box for member stars, where we take sources within 0.1 mag in color on either side of the isochrone at the tip of the red giant branch, and let the width of the box grow linearly to 0.3 mag by $i_0 = 24$\,mag. The slight stellar overdensity of isochrone-selected stars provided sufficient evidence to warrant further investigation.

\begin{figure*}
    \centering
    \includegraphics[width=0.9\linewidth]{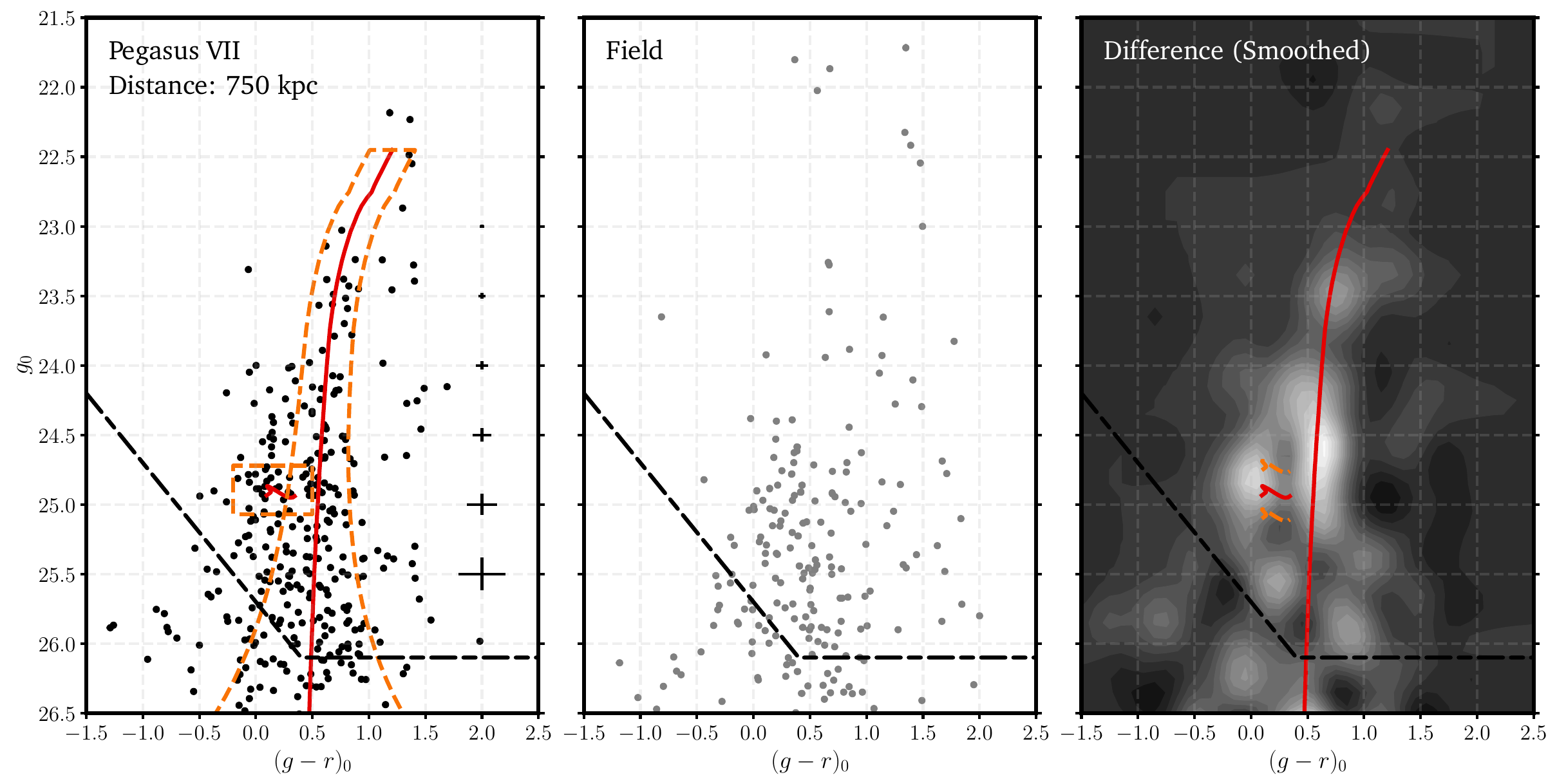}
    \caption{
    {\it Left}: CMD of stellar sources (black markers) within 2 half-light radii of the \peg\ centroid. An old (10\,Gyr), metal-poor ([Fe/H] = $-2$) PARSEC isochrone is overlaid in red, shifted to a distance of 750\,kpc. The 5$\sigma$ point source depths ($g \sim 26.1$ and $r \sim 25.7$\,mag) are indicated with a dashed black line. Black crosses show the median color and magnitude errors as a function of magnitude. Dashed orange outlines have been drawn to delineate where putative member stars of \peg\ fall on the CMD, as described in Section \ref{subsec:stardist}.
    {\it Centre}: This CMD shows sources which fall in an elliptical annulus whose inner radius is 3 half-light radii and has an area equal to the inner 2 half-light radius region, demonstrating the non-member source contamination present in the left panel.
    {\it Right}: Smoothed Hess difference plot between on-target and field CMDs. Light regions indicate a high density of \peg\ stars compared to field stars, while dark regions show deficits. The adopted isochrone is shown in red and we also shift the distance of the HB by 60\,kpc (orange dashed lines) to indicate how the distance uncertainty is inferred from the difference CMD.
    }
    \label{fig:peg7-cmd}
\end{figure*}

\subsection{Follow-up Imaging}

We were awarded time to further investigate this candidate detection with imaging using both the Canada-France-Hawaii Telescope (CFHT) and the Gemini Telescope.
We used the MegaCam wide-field camera \citep{Boulade2003} on CFHT through a Director's Discretionary Time proposal (program ID 23bd08, PI Smith). Observations were taken on 2023-10-24 and consisted of $3 \times 550$\,s with the $g$-band filter and $3 \times 510$\,s $r$-band filter, coming to a total program time of 1 hour.
At the Gemini North Telescope, we used the imaging mode of the Gemini Multi-Object Spectrograph (GMOS) through the use of the Fast Turnaround (FT) program (program ID GN-2023B-FT-202, PI Smith). Observations were taken on 2023-10-7 and consisted of $3 \times 440$\,s for both the $g$ and $r$ filters on GMOS for a total of 1320\,s of exposure time in each band. 
The Gemini photometry is deeper than the photometry derived from the CFHT imaging, so we use the Gemini follow-up data to confirm the reality of \peg\ and investigate its physical properties. We now describe the Gemini data reduction and photometry in further detail.


We reduce all individual GMOS images following the \textsc{Separate CCDs} recipe described in the \textsc{DRAGONS} \citep{Labrie23} documentation to preserve the different color responses of the three GMOS CCDs to better enable precision photometry. Additionally, we follow the same \textsc{DRAGONS} routine to reduce a fully stacked $g$-band image and a fully stacked $r$ band image.

To perform photometry, we first use \textsc{SExtractor} \citep{Bertin1996} to identify sources in the stacked $g$ image. 
We then measure the diameter of the seeing disk in each individual image using $\sim 30$ bright, isolated stellar sources. Using the source list from the $g$ image, we conduct forced aperture photometry on each individual image with an aperture whose diameter is 4$\times$ the seeing for that image, resulting in three measurements of the aperture photometry, per band. We tested apertures of different sizes and found 4$\times$ the seeing gives a fair balance between capturing most of the light without including too much background. The final magnitude is then taken as the median of three measurements and the final uncertainty as the quadrature sum of three uncertainties. For sources that fall on chip gaps or off-frame due to dithered frames, only on-frame measurements are combined for the final photometric catalog.

We validate the Gemini photometry derived here by comparing with UNIONS $gr$ photometry in the same region. The UNIONS $g$-band coverage is patchy in the region, so we cross-match UNIONS-$g$ with Gemini, and UNIONS-$r$ with Gemini independently. In Figure \ref{fig:offsets} we show the difference between UNIONS and Gemini photometry as a function of UNIONS photometry, given that the UNIONS photometry has been tested and validated on a large scale (Gwyn et al. 2024, in preparation).
In both the $g$ and $r$ bands, there are no trends with magnitude, though there is a clear constant offset. For the $g$-band, we measure an offset of 0.09\,mag using all sources between 19.5 and 22.5\,mag in UNIONS-$g$. For the $r$-band, we measure an offset of 0.07\,mag using all sources between 20 and 23\,mag in UNIONS-$r$. These constant offsets are applied to the Gemini aperture photometry.

We apply two cuts to select stars morphologically. We measure aperture photometry in $1 \times$ and $2 \times$ the seeing apertures on the stacks and take the difference of these magnitudes to obtain a measure of growth in the light curve. We identify the stellar locus and select all sources within 0.1\,mag at $g < 23$. The width of the stellar locus grows at faint magnitudes as it intersects with the cloud of extended sources, so we allow the stellar locus selection curve to grow to select sources within 0.25\,mag, which is roughly the 95\% percentile width of the locus at $g = 25$, and then continue constantly at that value to fainter sources. 
Additionally, there are a few regions at the edge of the images with obvious artifacts which are detected as a multitude of closely spaced sources. We experimented with different \textsc{SExtractor} parameters and found that taking the ratio of the semi-major and semi-minor axes to be less than 2 (essentially a circularity cut) helped to cull many of these unreliable sources.  

Using all sources that pass our stellar cuts, we measure the 5$\sigma$ median point source depth to be 26.1 and 25.7\,mag in $g$ and $r$ respectively. We again apply extinction corrections using $E(B-V)$ values from \citet{Schlegel98} assuming the conversion factors of \citet{Schlafly11}.

\subsection{Showing \peg\ to be a Real System}

The CMD based on our deeper Gemini data demonstrates that \peg\ is indeed a real stellar system.
Figure \ref{fig:peg7-cmd} shows the CMDs of extinction corrected $gr$ photometry for all stars within 2 elliptical half-light radii of the \peg\ centroid in the left panel. The central panel is a CMD of the field, specifically an equal-area ellipse whose inner radius is 3 half-light radii, to demonstrate source contamination in the left panel. The half-light radius and other structural parameters are computed in Section \ref{subsec:stardist}. The right panel of Figure \ref{fig:peg7-cmd} then shows the smoothed Hess difference plots between these two CMDs. Here, bright regions indicate high stellar densities, revealing a coherent stellar population. An excess of stars is visible between $0.2 \leq (g-r)_0 \leq 0.8$\,mag over a large range in $g_0$, which is consistent with a red giant branch (RGB), while an additional excess of bluer stars is prominent between $24.5 \leq g_0 \leq 25.0$\,mag, which is consistent with a horizontal branch (HB). Assuming the bluer overdensity is the HB at $g_0 \sim 24.85$\,mag, we get an initial distance estimate of roughly 700\,kpc. Stars consistent with the RGB and HB (Selection detailed in Section \ref{subsec:stardist}) are shown to be centrally concentrated in Figure \ref{fig:peg7-spatial}, further strengthening the detection of this faint stellar system.

\section{Results} \label{sec:res}

\subsection{Age, Metallicity, \& Distance}\label{subsec:basic}

\begin{figure*}
    \centering
    \includegraphics[width=0.45\linewidth]{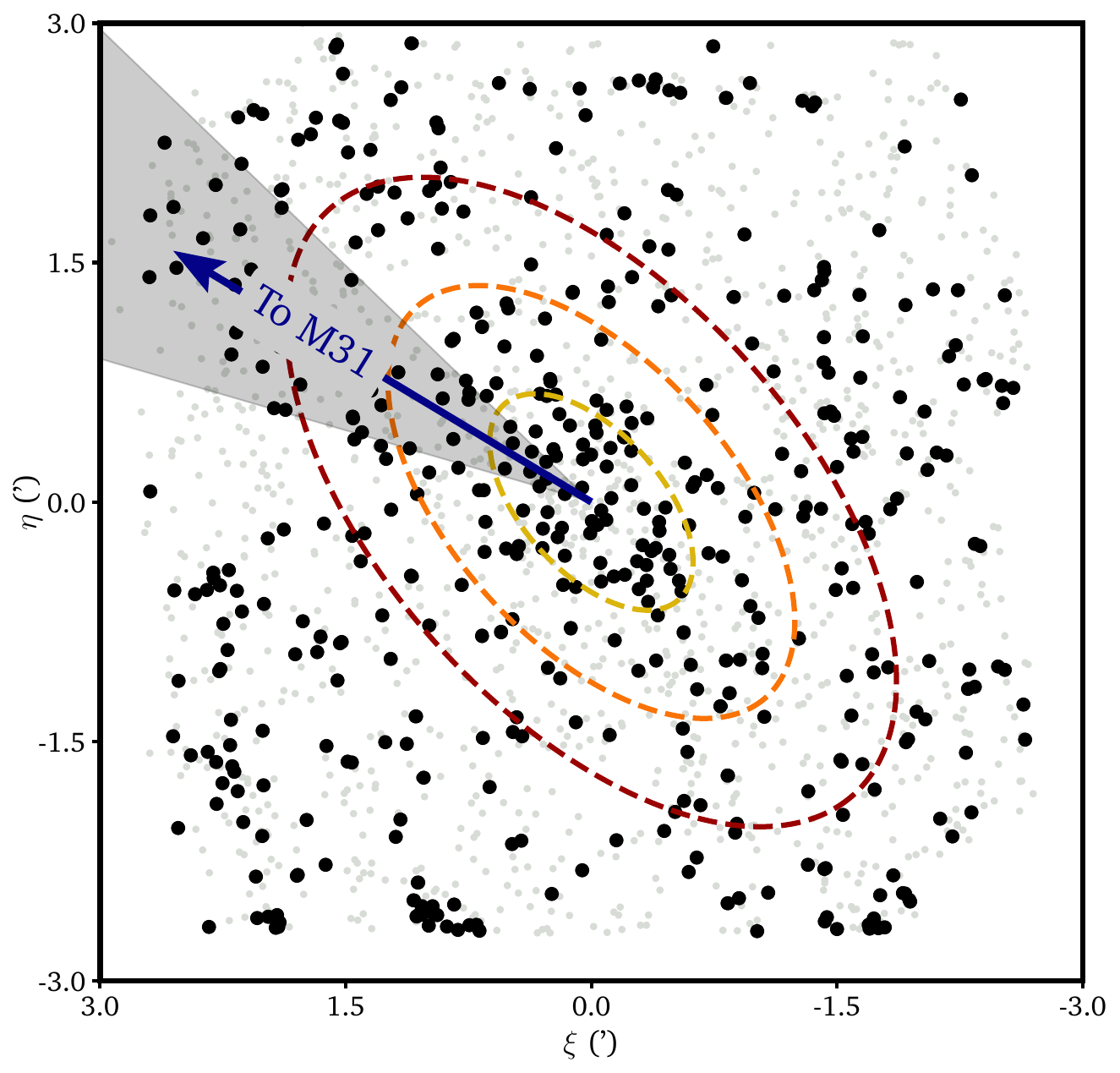}
    \includegraphics[width=0.45\linewidth]{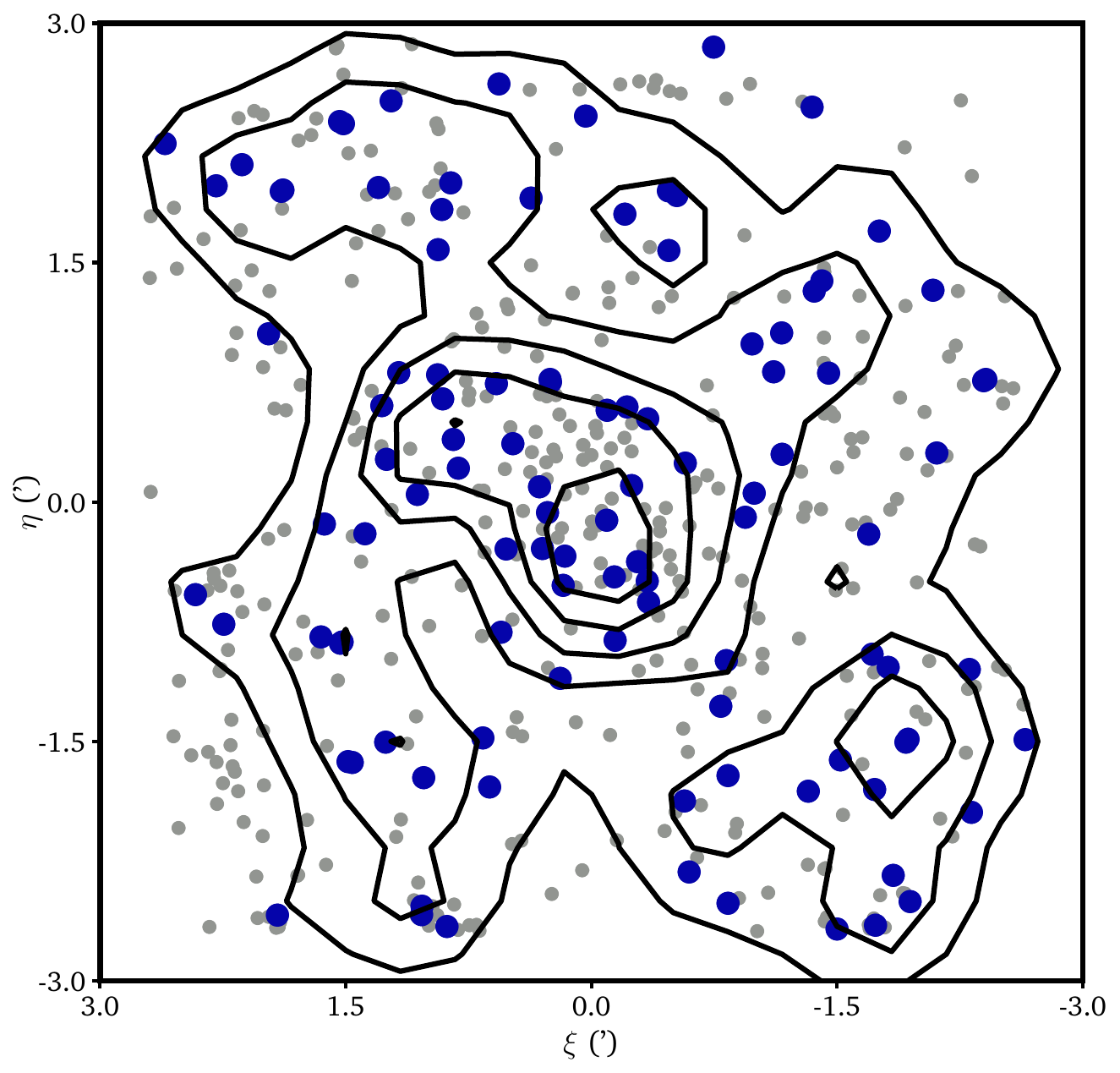}
    \caption{
    {\it Left}: Tangent plane-projected spatial distribution of cleaned stellar sources detected in Gemini-N/GMOS imaging about \peg. Black stars fall within the isochrone membership selection box defined in the left panel of Figure \ref{fig:peg7-cmd} while grey stars fall outside of this selection box. Concentric ellipses in yellow, orange, and red show the 1, 2, \& 3\,$\times$\,\rh\ iso-density contours, respectively, of the best-fit stellar distribution model. The blue arrow indicates the tangent plane-projected direction towards M31 and the grey wedge shows the 1$\sigma$ uncertainties on the projected direction towards M31 stemming from the position angle of the major axis of \peg. The direction towards M31 is 18$^{+12}_{-9}$\,$\deg$\ from the position angle of the major axis.
    {\it Right}: Here we plot in large blue markers the tangent plane-projected spatial distribution of all suspected horizontal branch stars (defined by the HB selection box in the left panel of Figure \ref{fig:peg7-cmd}). We overlay black contours showing the smoothed density distribution of these same HB stars. We also plot small grey markers for all candidate member stars (same as black point in left panel of this figure), demonstrating how the HB distribution traces the same overdensity as the rest of the member stars.
    }
    \label{fig:peg7-spatial}
\end{figure*}

We use PARSEC isochrones to further constrain the heliocentric distance of \peg\ while simultaneously obtaining an estimate for the age and metallicity for the stellar population. We visually inspect a grid of isochrones spanning [Fe/H] of $-2.2$ to $-1.4$\,dex in steps of 0.1\,dex, and ages ($\tau$) of 7 to 13\,Gyr in steps of 1\,Gyr. These isochrones were overlaid on the Hess diagram and shifted to various distances around 700\,kpc. The shape of the RGB provides a mild metallicity constraint as more metal-rich isochrones are redder at fixed age. Additionally, we require the mean HB magnitude in the theoretical isochrones to be within roughly 0.2\,mag of the measured HB ($g_0 \sim 24.85$\,mag). There are no suitable isochrones for distances closer than 690\,kpc or further than 810\,kpc. In Figure \ref{fig:peg7-cmd}, we note that two stars fall near to the tip of the RGB, at a color of $(g-r)_0 \sim 1.4$. If all sources in the deep imaging were to be plotted on the CMD, there is a prominent feature at this color over a large magnitude range which we suspect to be the faint end of the Galactic disk sequence. The field CMD has a few sources at the same color and magnitude, so we do not consider it likely that the two stars near the tip of the RGB are members of the system and exclude them from our distance determination. Within this distance range, all isochrones more metal-rich than $-1.7$\,dex fail to trace the upper portion of the RGB. It would be unusual for the dominant stellar population of a faint M31 dwarf galaxy to be younger than $\sim10$\,Gyr \citep{Savino2023}, but it is very difficult to constrain the age of the stellar population without reaching the Oldest Main-Sequence Turn-Off (oMSTO), so we do not constrain the age any further than the initial range that was considered. 
This gives a likely range for [Fe/H] of $-2.2$ to $-1.8$\,dex, $\tau$ within 7 to 13\,Gyr, and heliocentric distance within 690 to 810\,kpc. We note that PARSEC isochrones are not computed below [Fe/H] = $-2.2$, so the metallicity is unconstrained beyond this point.
We adopt a representative isochrone for the remainder of this work as the midpoint of all parameters, namely [Fe/H] = $-2$ and $\tau = 10$\,Gyr at a distance of $750$\,kpc. In Table \ref{tab:props}, we list each parameter as the midpoint with errorbars that indicate the possible range from this qualitative analysis.

\begin{deluxetable}{llc}
\tabletypesize{\footnotesize}
\tablecaption{Measured and derived properties for Pegasus VII \label{tab:props}}

\tablehead{
    Property & Description & Value
}

\startdata
$\alpha_{J2000}$ & Right Ascension & 23$h$ 1$m$ 49.3\,$\pm$\,0.5$s$ \\
$\delta_{J2000}$ & Declination & +32\textdegree\ 5\arcmin\ 52\,$\pm$\,6\arcsec \\
$r_{\text{h,ang}}$ & Angular Half-Light Radius & 0.8$^{+0.2}_{-0.1}$\,\arcmin \\
$r_{\text{h,phys}}$ & Physical Half-Light Radius & 177$^{+36}_{-34}$\,pc \\
$\epsilon$ & Ellipticity & 0.5$^{+0.1}_{-0.2}$ \\
$\theta$ & Position Angle & 40$^{+12}_{-9}$\,deg \\
N$^*$ & Number of Stars\tablenotemark{a}  & 82$^{+15}_{-14}$ \\
$\tau$ & Age (Isochrone) & $10$\,$\pm$\,3\,Gyr\tablenotemark{b}\\
\met & Metallicity (Isochrone) & $-2$\,$\pm$\,0.2\,dex\tablenotemark{b}\\
$D_{\odot}$ & Heliocentric Distance & 750\,$\pm$\,60\,kpc\tablenotemark{b}\\
$D_{\text{M31}}$ & 3D Distance to M31 & 331$^{+15}_{-4}$\,kpc\\
M$_{*}$ & Total Stellar Mass & $2.6^{+0.6}_{-0.5}$\,$\times$\,10$^4$\,\Msun \\
$M_V$ & Absolute $V$-band Magnitude & $-5.7$\,$\pm$\,0.2\,mag \\
$\mu_{0}$ & Central Surface Brightness & $27.3$\,$\pm$\,0.5\,mag\,arcsec$^{-2}$ \\
\enddata

\tablenotetext{a}{Number of stars down to $g = 25.6$\,mag, 0.5\,mag brighter than the 5$\sigma$ point source depth in $g$.}
\tablenotetext{b}{Age, metallicity, and distance are all estimated qualitatively by visual inspection of isochrones. Please note that the values should be taken as the midpoint and range of possible parameters, rather than a fit to the data.}

\end{deluxetable}

\subsection{Stellar Distribution}\label{subsec:stardist}

To assess the stellar distribution of \peg\ we first define a membership selection box on the CMD. To select RGB stars, we define a region about the adopted isochrone that extends 0.2\,mag both bluer and redder at all magnitudes, which accounts both for uncertainties and intrinsic variations in age and metallicity. We also add twice the median photometric error as a function of magnitude to the intrinsic box width, to account for larger uncertainties at fainter magnitudes.
Additionally, we draw a box around the nominal HB whose color ranges from $-0.2 \leq (g-r)_0 \leq 0.5$\,mag, and whose height ranges from $24.7 \leq g_0 \leq 25.1$\,mag. The magnitude range is set by finding the mean $g_0$ of the HB ($\langle g_{0, \text{HB}} \rangle$) at 750\,kpc, then setting the upper and lower bounds as $\langle g_{0, \text{HB}} \rangle$ shifted to $\pm 60$\,kpc, the adopted uncertainty on the distance estimate. This selection box is drawn on the target CMD (left panel) in Figure \ref{fig:peg7-cmd}. 

We now estimate the structural parameters of \peg\ by only considering the spatial distribution of stars which fall within the CMD selection box. We follow the methodology of \citet{Martin08, Martin16-pds} to measure these parameters by constructing a stellar distribution model composed of an elliptical, exponential radial surface density profile and a constant density background. The model is described by six free parameters: the coordinates of the profile centroid ($x_0, y_0$), the ellipticity $\epsilon$ (defined as $\epsilon = 1 - b/a$ where $b/a$ is the minor-to-major-axis ratio of the model), the position angle of the major axis $\theta$, (defined East of North), the half-light radius (which is the length of the semi-major axis \rh), and the number of stars $N^*$ in the system; each best-fit parameter is estimated using a Markov Chain Monte Carlo method to sample the posterior distribution.
A full description of this procedure is given in \citet{Smith24}, so we refer the reader there for more details. 

When selecting the list of input stars, we also imposed a magnitude limit of $g_0 \sim 25.6$\,mag (0.5\,mag brighter than the 5$\sigma$ depth in $g$) to help mitigate for stellar incompleteness. 
We experimented with different limits between 25.1 and 26.1\,mag and found them to produce resultant best-fit parameters that are self-consistent within uncertainties.

The median values for each best-fit parameter are listed in Table \ref{tab:props} where uncertainties are given as the 16th and 84th quantiles of each posterior distribution. The angular half-light radius is measured to be 0.8$^{+0.2}_{-0.1}$\,\arcmin\ which, when combined with the distance estimate, gives a physical half-light radius of 177$^{+36}_{-34}$\,pc.
The 1, 2, \& 3\,$\times$\,\rh\ iso-density contours of the best-fitting structural parameters are over-plotted on the spatial distribution of sources in the left panel of Figure \ref{fig:peg7-spatial}. The right panel of Figure \ref{fig:peg7-spatial} shows the distribution of HB stars (selected using the box in the left panel of Figure \ref{fig:peg7-cmd}). The clustering of selected sources in the centre of the plot suggests that we are selecting {\it bona fide} HB stars, though there is still contamination in the box likely stemming from blue background galaxies that pass our point-source cuts.

\subsection{Mass \& Magnitude}\label{subsec:totals}

We estimate the total stellar mass ($\text{M}_{*}$) and total absolute $V$-band magnitude ($M_V$) of \peg\ following an identical procedure to that of \citet{Smith24}, which is in turn based on \citet{Martin16-pds}. We refer the reader to \citet{Smith24} for a full description of the synthetic stellar populations generated. Critical to the estimation performed here, we use a heliocentric distance of $750$\,$\pm$\,$60$\,kpc and set the total number of stars brighter than $g_0 = 25.6$\,mag to be 82\,$\pm$\,15. Uncertainties are assumed to be drawn from Gaussian distributions when propagated through to the final magnitude and mass.

We generate 10,000 synthetic stellar populations and measure the total mass and magnitude of each. The final derived quantities for \peg\ are then taken as the median value of each distribution with 16th and 84th quantiles as the 1$\sigma$ lower and upper bounds. We measure $\text{M}_{*} = 2.6^{+0.6}_{-0.5}$\,$\times$\,10$^4$\,\Msun\ and $M_V = -5.7$\,$\pm$\,0.2\,mag. We also measure the central surface brightness ($\mu_{0}$) to be $27.3$\,$\pm$\,0.5\,mag\,arcsec$^{-2}$.

\begin{figure*}
    \centering
    \includegraphics[width=\linewidth]{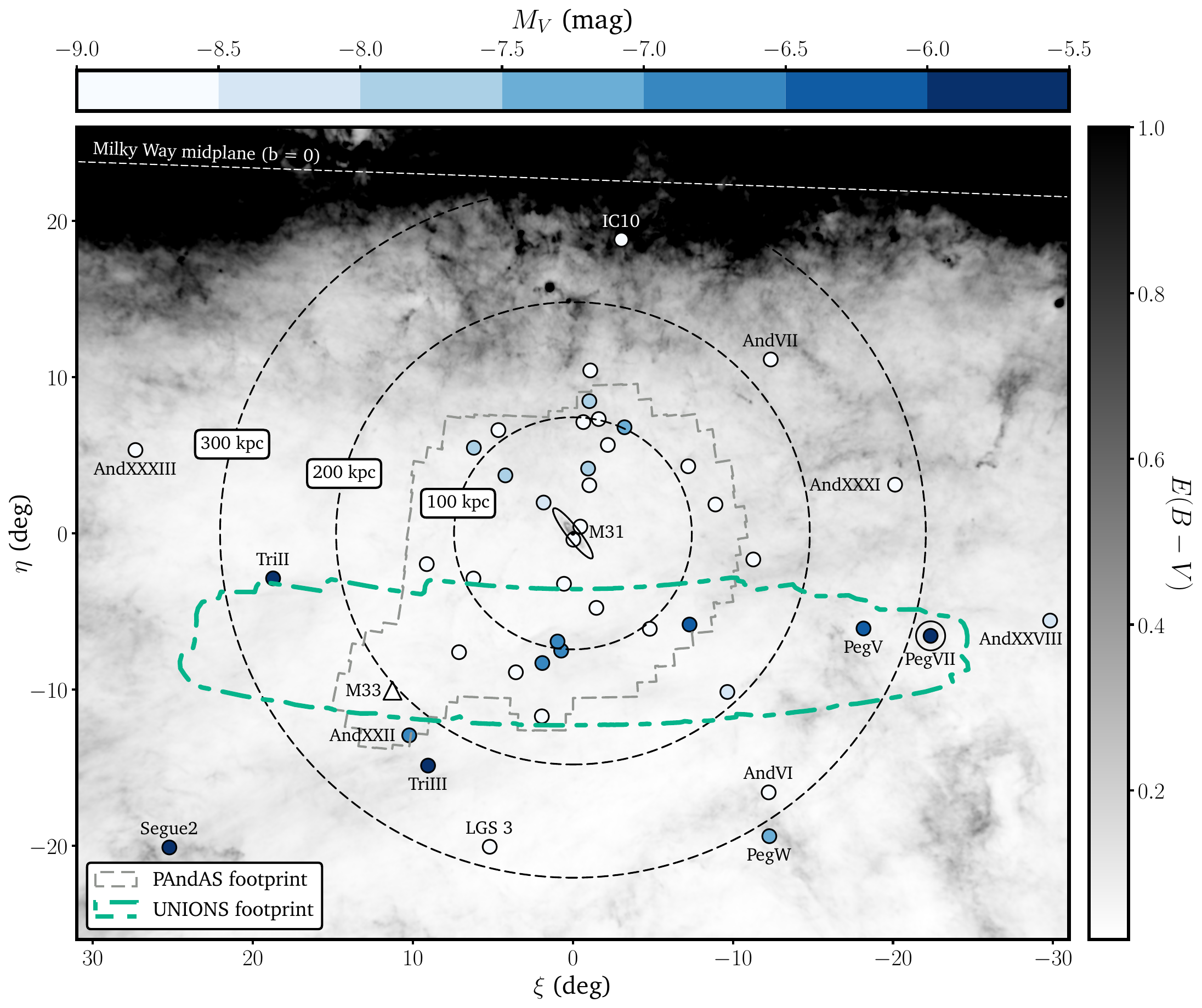}
    \caption{Tangent plane projection centered on M31, which is shown as the central black ellipse. 
    Known galaxies in the vicinity of M31 are shown as black circles where their filled shade of blue indicates total absolute $V$-band magnitude in the range indicated in the color bar. The position of M33 (Triangulum) is marked with a white triangle. The three concentric, dashed black circles indicate projected radii of 100, 200, \& 300\,kpc at the distance of M31. The dashed grey outline demarcates the PAndAS footprint while the UNIONS footprint in the south galactic cap is shown with a thick dashed green line, as noted in the legend. 
    \peg\ is indicated with a label as well as a large black circle and lies in front of the tangent plane, at a total three-dimensional distance of $\sim$330\,kpc from M31.
    All known galaxies in this field of view with projected M31 separations greater than 200\,kpc are labelled.
    The greyscale background shows the dust maps from \citet{Schlegel98} where the color bar has a maximum of $E(B-V) = 1$ to show the features around the M31 satellites, though the top band reaches much greater extinctions towards the galactic midplane (indicated by the dashed white line). Triangulum II and Segue 2 are Milky Way satellites in the foreground of this field of view.}
    \label{fig:m31-field}
\end{figure*}

\section{Discussion and Conclusions} \label{sec:disc}

The faintest previously known galaxies in the M31 halo have been measured to be between $M_V = -6$ and $M_V = -6.5$ \citep{Martin16-pds, Savino2022}. The measurement of $M_V = -5.7$\,$\pm$\,0.2\,mag for \peg\ using these Gemini data place it firmly in the ultra-faint dwarf (UFD) galaxy regime \citep[$M_V > -7.7$,][]{Bullock17} and make it the faintest known M31 dwarf galaxy satellite discovered to date, though deeper data will be necessary to compare with the total absolute magnitudes derived for other M31 dwarfs from HST observations \citep{Savino2022}. With a physical half-light radius of $\sim$177\,pc, \peg\ is roughly 5x larger than the most extended M31 globular clusters \citep[\rh\,$\sim$\,30\,pc;][]{Huxor2014}, making its status as a dwarf galaxy highly likely.
Comparable dwarf galaxy satellites of the Milky Way in terms of size and magnitude are Bo\"otes I ($M_V = -6.0$, $r_{\text{h, phys}} = 202$\,pc), Hercules ($M_V = -5.8$, $r_{\text{h, phys}} = 224$\,pc) and Ursa Major I ($M_V = -5.1$, $r_{\text{h, phys}} = 229$\,pc) \citep[][]{Munoz2018}. However, unlike these galaxies, \peg\ finds itself at a rather large three-dimensional separation from its host, at a distance of 331$^{+15}_{-4}$\,kpc (where an M31 distance of 770\,kpc is adopted), beyond the nominal virial radius of the M31 dark matter halo \citep[$\sim$240 to 300\,kpc from][]{BlanaDiaz2018, Kafle2018, Putman2021}.
Dwarf galaxies in the vicinity of M31 are shown in Figure \ref{fig:m31-field}, with the PAndAS and UNIONS footprints overlaid, along with projected 100, 200, \& 300\,kpc radii. This raises the question as to whether or not \peg\ has previously passed within the virial radius of M31.


Largely speaking, there are two possible interaction histories for \peg: (1) this galaxy has previously interacted with M31, or (2) \peg\ is on first infall.
We do not have a line-of-sight velocity, nor do we have proper motions, so determining the orbit of \peg\ is not possible at present. 
However, the stellar distribution of \peg\ may provide an opportunity to investigate this question.
\citet{Pace2022} studied the orbits of the Milky Way satellite group and compared the direction of projected orbital motion with the projected ellipticity of the stellar distribution. Their findings indicated a preference for dwarf galaxies with large ellipticities ($\epsilon > 0.4$) to be aligned with their orbital motion. 
Numerical simulations by \citet{Pereira2008} suggest that after infall (crossing the host virial radius), satellite galaxies in clusters show preferential radial alignment with the direction towards their hosts, with no preferential alignment present prior to infall. Alignment evolves over subsequent orbits, but is on average strongest at apocentre. A similar result is obtained from the Aquarius Project by \citet{Barber2015} who find preliminary evidence of radial alignment between M31 and many of its hosts. They suggest that radial alignment is an indicator that multiple orbits have been completed and that reaching this state necessitates significant tidal stripping. 

\peg\ has an ellipticity of $\epsilon = 0.5^{+0.1}_{-0.2}$ and this projected elongation is aligned within 18$^{+12}_{-9}$\,deg of the projected direction towards M31, as shown in Figure \ref{fig:peg7-spatial}. If it has had a previous interaction, it is likely out near apocentre or even moving beyond the virial radius \citep[if it is a backsplash galaxy; e.g.,][]{SantosSantos2023}. It is therefore feasible that the elongation of \peg\ indicates both the direction of motion (either away or towards M31) and that the source of this elongation is a previous tidal interaction with the M31 gravitational potential. 

If \peg\ is on first infall then, at a current separation of $\sim$330\,kpc, it is just about to cross the virial radius of M31 and has likely been isolated up until this point. The Solitary Local (Solo) Dwarf Galaxy Survey \citep{Higgs21} imaged twelve northern isolated Local Group dwarfs spanning $M_V$ of $-8$ to $-17$, and compared their properties to the satellite populations of the MW and M31. Solo dwarf galaxies at large distances ($> 400$\,kpc) from both the MW and M31 have preferentially large ellipticities ($\epsilon \gtrsim 0.3$). While this sample does not include any galaxies as faint as \peg\ ($M_V \sim -5.7$), this result suggests that isolated systems can exhibit elongation in the absence of tidal interactions.
Recent theoretical work from \citet{Goater2024} in the Engineering Dwarfs at Galaxy Formation's Edge (EDGE) cosmological simulation suite offer a possible, natal origin for ellipticity in low mass dwarfs. They studied isolated dwarf galaxies with stellar masses of $0.07$-$2$\,$\times$10$^{6}$\,\Msun\ and found them to exhibit a wide range in projected ellipticities ($0.03 < \epsilon < 0.85$) which can be causally linked to their formation timescales. Specifically, they find that dwarfs which form the bulk of their stars from {\it in situ} gas will have rounder, more compact stellar systems, while dwarfs that accrue {\it ex situ} stars through a late-time dry accretion merger will be more elliptical. Therefore, the measured elongation of \peg\ could indicate that a dwarf-dwarf merger is responsible for the bulk of its stellar mass assembly. 

The findings that elongation can stem from multiple processes highlights that no robust claim can be made about the evolutionary history of \peg\ from these photometric data alone. A full orbital analysis will be necessary to understand whether \peg\ has previously interacted with M31 or is on first infall. Stellar spectroscopy and long temporal baseline astrometry will be necessary to establish a line-of-sight velocity and proper motion, respectively. Deep, space-based imaging reaching below the oMSTO will allow for the calculation of the star formation history (SFH), while also setting a first epoch for future proper motion measurements.
Combining the orbit and SFH could provide detailed insight into how reionization, possible interactions with M31, or a dwarf-dwarf merger may have shaped the evolution of \peg, as has been done for other Local Group dwarfs \citep[e.g. Leo I;][]{RuizLara2021, Bennet2024}.
Additionally, follow-up observations to investigate the presence of gas in \peg\ may indicate whether ram pressure stripping in the M31 halo has significantly impacted the dwarf galaxy \citep[e.g.,][]{Spekkens2014, Putman2021}, where a dearth of gas is a clear prediction for galaxies that have completed 1 or more orbits of their host \citep[e.g.,][]{Barber2015}. 
If \peg\ does not show signs of major interactions with a host in its orbital history or gas content, it could provide a window into the evolutionary pressures faced by a low mass field dwarf in the Local Group.


Under the assumption that \peg\ is a satellite of M31, its discovery in UNIONS also offers an observational perspective on the M31 dwarf galaxy population. With a heliocentric distance of 750\,$\pm$\,60\,kpc and at a physical separation of 331$^{+15}_{-4}$\,kpc, \peg\ lies in front of the M31 midplane, adding to the well documented lopsidedness of the M31 sub-group \citep{McConnachie2006, Conn2013, Savino2022}. \citet{Doliva2023} find this satellite distribution anisotropy to be highly statistically significant when considering completeness limits in the PAndAS footprint.
\citet{Doliva2023} also predict that between 60 and 100 satellites brighter than $M_V < -5.5$ remain to be discovered out to a projected distance of 300\,kpc. Beyond 300\,kpc, the APOSTLE\footnote{A Project Of Simulating The Local Environment} cosmological hydrodynamics simulations \citep{Fattahi2016, Sawala2016} predict that there are $\sim 50$ dwarf galaxies with $M_* \gtrsim 10^5$\,\Msun\ missing from the Local Group galaxy inventory, with a significant fraction of those expected to be discovered in the direction of M31 but outside of its virial radius \citep{Fattahi2020, SantosSantos2024}.
The discovery of \peg\ complements both the empirical and theoretical claim that a wealth of dwarf galaxy satellites remain undetected towards M31. 

Excluding the Milky Way satellites Tri II and Segue 2, thirteen galaxies lie more than 200\,kpc from M31 in the field of view presented in Figure \ref{fig:m31-field}, where And XXVII, And XXXI (or Lac I), and And XXXIII (or Per I) have been confirmed as bound to M31 through radial velocity measurements \citep{Slater2015, Martin2014-three}. 
The faintest of these distant systems (Peg V \citep{Collins22}, Peg W \citep{McQuinn23-PegW}, Tri III \citep{Collins23} and now \peg) have been recently discovered in the southern half of the M31 halo in deep, wide-field surveys (UNIONS and DESI Legacy Surveys). 
The northern side of the M31 region is shrouded in significant dust extinction, as shown in Figure \ref{fig:m31-field}, but the clear presence of systems with $M_V < -5.5$ beyond 200\,kpc highlights the need for dedicated imaging that pushes towards low galactic latitudes in order to accurately measure the completeness and spatial distribution of M31 satellite galaxies.
The extent of UNIONS over a more distant region of the M31 halo than PAndAS provides the opportunity to hunt for more outer M31 satellites and potential M33 satellites via matched-filter methods complemented by visual inspection through both current and future UNIONS data releases.


\section*{Acknowledgments}

We thank the reviewer Nelson Caldwell for their keen and careful comments that helped improve the quality of this manuscript.

We would like to respectfully acknowledge the L\textschwa\textvbaraccent {k}$^{\rm w}$\textschwa\textipa{\ng}\textschwa n Peoples on whose traditional territory the University of Victoria stands, where the majority of this work was carried out. We strive to honour the Songhees, Esquimalt and $\underline{\text{W}}\acute{\text{S}}$ANE$\acute{\text{C}}$ peoples who were the first astronomers of this land and whose continued stewardship is crucial to its preservation.

As stated in individual acknowledgments below, data collection for this work was conducted at several observing sites atop Maunakea. Therefore, the authors wish to recognize and acknowledge the very significant cultural role and reverence that the summit of Maunakea has always had within the Native Hawaiian community. We are most fortunate to have the opportunity to conduct observations from this mountain.

This work is based on data obtained as part of the Canada-France Imaging Survey, a CFHT large program of the National Research Council of Canada and the French Centre National de la Recherche Scientifique. Based on observations obtained with MegaPrime/MegaCam, a joint project of CFHT and CEA Saclay, at the Canada-France-Hawaii Telescope (CFHT) which is operated by the National Research Council (NRC) of Canada, the Institut National des Science de l’Univers (INSU) of the Centre National de la Recherche Scientifique (CNRS) of France, and the University of Hawaii. This research used the facilities of the Canadian Astronomy Data Centre operated by the National Research Council of Canada with the support of the Canadian Space Agency. This research is based in part on data collected at Subaru Telescope, which is operated by the National Astronomical Observatory of Japan. 
Pan-STARRS is a project of the Institute for Astronomy of the University of Hawaii, and is supported by the NASA SSO Near Earth Observation Program under grants 80NSSC18K0971, NNX14AM74G, NNX12AR65G, NNX13AQ47G, NNX08AR22G, YORPD20\_2-0014 and by the State of Hawaii. 

This work is based on observations obtained at the international Gemini Observatory and processed using DRAGONS (Data Reduction for Astronomy from Gemini Observatory North and South). The international Gemini Observatory is a program of NSF NOIRLab, which is managed by the Association of Universities for Research in Astronomy (AURA) under a cooperative agreement with the U.S. National Science Foundation on behalf of the Gemini Observatory partnership: the U.S. National Science Foundation (United States), National Research Council (Canada), Agencia Nacional de Investigaci\'{o}n y Desarrollo (Chile), Ministerio de Ciencia, Tecnolog\'{i}a e Innovaci\'{o}n (Argentina), Minist\'{e}rio da Ci\^{e}ncia, Tecnologia, Inova\c{c}\~{o}es e Comunica\c{c}\~{o}es (Brazil), and Korea Astronomy and Space Science Institute (Republic of Korea).

\vspace{5mm}
\facilities{CFHT, Subaru, Gemini:Gillett}


\software{\texttt{astropy} \citep{Astropy13, Astropy18, Astropy22}, \texttt{emcee} \citep{Foreman13}, \texttt{numpy} \citep{Numpy20}, \texttt{scipy} \citep{Scipy20}}

\bibliography{ref}{}
\bibliographystyle{aasjournal}



\end{document}